\documentclass[10pt,aps,prd,floatfix,notitlepage,nofootinbib]{revtex4-1}
\usepackage[utf8]{inputenc}
\usepackage{multirow}
\usepackage{latexsym}
\usepackage{epsfig,graphics,subfig,float}
\usepackage{epstopdf}

\usepackage{tabularx,booktabs,caption}

\usepackage{amssymb}
\usepackage{verbatim}
\usepackage{multirow}
\usepackage{color}
\usepackage{tikz}
\usepackage{amsmath,amssymb,amsfonts}

\usepackage{tikz}
\usepackage{bm}
\usetikzlibrary{matrix}

\begin{document}

\title{Linear potentials in galaxy halos by Asymmetric Wormholes}
\author{Sebastian Bahamonde}
\email{sebastian.beltran.14@ucl.ac.uk, sbahamondebeltran@gmail.com}
\affiliation{Department of Mathematics, University College London,
	Gower Street, London, WC1E 6BT, United Kingdom}
    \affiliation{   School of Mathematics and Physics, University of Lincoln.
   	Brayford Pool, Lincoln, LN6 7TS, United Kingdom}
    \affiliation{University of Cambridge, Cavendish Laboratory, JJ Thomson Avenue, Cambridge CB3 0HE, United Kingdom}

\author{David Benisty}
\email{benidav@post.bgu.ac.il}
\affiliation{Physics Department, Ben-Gurion University of the Negev, Beer-Sheva 84105, Israel}
\affiliation{Fachbereich Physik der Johann Wolfgang Goethe Universit\"{a}t, Campus Riedberg, Frankfurt am Main, Germany}
\affiliation{Frankfurt Institute for Advanced Studies, Giersch Science Center, Campus Riedberg, Frankfurt am Main, Germany}

\author{Eduardo I. Guendelman}
\email{guendel@bgu.ac.il}
\affiliation{Physics Department, Ben-Gurion University of the Negev, Beer-Sheva 84105, Israel}
\affiliation{Frankfurt Institute for Advanced Studies, Giersch Science Center, Campus Riedberg, Frankfurt am Main, Germany}
\affiliation{Bahamas Advanced Study Institute and Conferences, 4A Ocean Heights, Hill View Circle, Stella Maris, Long Island, The Bahamas}


\begin{abstract}
A spherically symmetric space-time solution for a diffusive two measures theory is studied. An asymmetric wormhole geometry is obtained where the metric coefficients have a linear term for galactic distances and the analysis of Mannheim and collaborators, can then be used to describe the galactic rotation curves. For cosmological distances, a de-Sitter space-time is realized. Centre of gravity coordinates for the wormhole is introduced which are the most suitable for the collective motion of a wormhole. The wormholes connect universes with different vacuum energy densities which may represent different universes in a ``landscape scenario". The metric coefficients depend on the asymmetric wormhole parameters. The coefficient of the linear potential is proportional to both the mass of the wormhole and the cosmological constant of the observed universe. Similar results are also expected in other theories like $k$-essence theories, that may support wormholes.
\end{abstract}

\date{\today}

\maketitle
	
\section{Introduction}\label{sec:0}
One of the most challenging questions in astrophysics is the mismatch between the measurements of the velocities of stars in galaxies, and the predictions for galaxy rotation curves from the standard general theory of relativity. This question led the astrophysicists arguing about the existence of dark matter. Other theorists instead have tried to modify General Relativity (GR) or the Newtonian laws. The most prevailing belief is that for explaining the galaxy rotation curves, the galaxy has to be soaked in a dark matter halo \cite{Rubin:1980zd,Rubin:1970zza,Kulchoakrungsun:2017ymx}. Regardless of this question, a theoretical spherically symmetric solution for GR, called a wormhole, where two different universes can be causally connected through a “wormhole throat” and where a physical traveller can go in principle from one universe to the other and be observed doing this from an observer located in one of the universes, gives a concept of a
non-trivial topological structure linking separate points in space-time~\cite{Lobo:2005us,Morris:1988cz,Sushkov:2005kj,Visser:1989kh}. This property of the space-time is different from black holes solutions, which may also connect two universes, but possess event horizons, so that the trajectory of the traveller cannot be followed by an external observer beyond the point where the observer crosses the horizon. 

The existence of a multi universe, each of them with different vacuum energy density, has been widely discussed in both string theory and in inflationary cosmology. One possibility is that these universes were initially connected in the early stages of evolution but are now totally disconnected. Another  possibility  that appears more interesting  is  that some of these universes  are now still connected through a wormhole and may be this connection leads to observable consequences. 

In this letter, we propose a modified theory of gravity, which for  spherically symmetric solutions produces asymmetric wormholes. The asymmetry is a consequence of the fact that in our case the wormholes  connect universes with different vacuum energy densities,  and therefore are necessarily asymmetric. For large distances the solution produces gravitational potentials that can be suitable for the explanation of galaxy rotation curves. The parameters that defines the asymmetry of the wormhole hole determine linear gravitational potentials and therefore could provide an explanation for the rotation curves in galaxy halos. It is interesting to note that the possibility that  the massive object detected at the centre of our galaxy is a wormhole rather than a black hole has been discussed together with some possible observational consequences related to the effect of this on the geodesics produced by this object~\cite{Zilong}. These effects are indeed even more acute in the case of the solutions discussed in this paper  due to the generation of the linear potentials. 
\subsection{Two Measures Theory}

Many modified theories of gravity have been formulated for explaining phenomena beyond GR. One example is the two-measures theory \cite{TMT1,TMT2,TMT3,TMT4,TMT5,TMT6,TMT7,TMT8,TMT9} where in addition to the regular measure of integration in the action $ \sqrt{-g} $, includes another measure of interaction which is also a density volume and a total derivative. In this case, one can use for constructing this measure 4 scalar fields $ \varphi_{a} $, where $ a=1,2,3,4 $. Then, we can define the density $ \Phi=\varepsilon^{\alpha\beta\gamma\delta}\varepsilon_{abcd}\partial_{\alpha}\varphi_{a}\partial_{\beta}\varphi_{b}\partial_{\gamma}\varphi_{c}\partial_{\delta}\varphi_{d} $, and then we can write an action that uses both of these densities:
\begin{equation}
        S=\int d^{4}x\Phi\mathcal{L}_{1}+\int d^{4}x\sqrt{-g}\mathcal{L}_{2}\,.
\end{equation}
As a consequence of the variation with respect to the scalar fields $ \varphi_{a} $, assuming that $ \mathcal{L}_{1} $
and $ \mathcal{L}_{2} $
are independent of the scalar fields $\varphi_{a}$, we obtain that
\begin{equation} \label{measure}
        A_{a}^{\alpha}\partial_{\alpha}\mathcal{L}_{1}=0\,,
\end{equation}
where $ A_{a}^{\alpha}=\varepsilon^{\alpha\beta\gamma\delta}\varepsilon_{abcd}\partial_{\beta}\varphi_{b}\partial_{\gamma}\varphi_{c}\partial_{\delta}\varphi_{d} $. Since $ \det[A_{a}^{\alpha}]\sim\Phi^{3} $, then for $ \Phi\neq0 $, (\ref{measure}) implies that $\mathcal{L}_{1}=M=const$. This result can be expressed as a covariant conservation of a stress energy momentum of the form $ T_{\left(\Phi\right)}^{\mu\nu}=\mathcal{L}_{1}g^{\mu\nu} $, and using the 2nd order formalism where the covariant derivative of $ g_{\mu\nu} $ is zero, we obtain that $ \nabla_{\mu}T_{\left(\Phi\right)}^{\mu\nu}=0 $ implying ${\partial_{\alpha}\mathcal{L}_{1}=0}$. This suggests the idea of generalising the two-measures theory by imposing the covariant conservation of a non-trivial kind of energy-momentum tensor, which we denote as $ T_{\left(\chi\right)}^{\mu\nu} $ \cite{G1}.
Therefore, we consider an action of the form
\begin{equation} \label{11}
	S=S_{\left(\chi\right)}+S_{\left(R\right)}=\int d^{4}x\sqrt{-g}\chi_{\mu;\nu}T_{\left(\chi\right)}^{\mu\nu}+\frac{1}{2\kappa^2}\int d^{4}x\sqrt{-g}R\,,
\end{equation}
where semicolon $;$ denotes covariant derivative, $\kappa^2=8\pi G$ and $ \chi_{\mu}$ is the dynamical vector field. If we assume $ T_{\left(\chi\right)}^{\mu\nu} $ to be independent of $  \chi_{\mu} $  and having $ \Gamma_{\mu\nu}^{\lambda} $ being defined as the Christoffel connection coefficients, then the variation with respect to $ \chi_{\mu} $ gives a covariant conservation: $ \nabla_{\mu}T_{\left(\chi\right)}^{\mu\nu}=0 $. A full phenomenology for using these theories is described in \cite{Guendelman:2009ck,Benisty:2016ybt}.

\subsection{Diffusive Energy theory from Action principle}
Calogero \cite{Calogero:2011re} proved that the diffusion equation in a curved space-time implies a non-conserved stress energy tensor $T^{\mu\nu}$, which has some current source $f^\mu$:
\begin{equation} \label{chitensor}
	\nabla_\nu T^{\mu\nu}=3\sigma f^\mu\,,
\end{equation}
where $\sigma$ is the diffusion coefficient of the fluid. This generalisation is Lorentz invariant and the current $ f^\mu$  is a time-like covariantly conserved vector field and its conservation tells us that the number of particles in this fluid is constant. This non-conservative stress energy tensor can emerge from variations in the action (\ref{11}), by replacing the dynamical time vector field for a gradient of a scalar field $\partial_\mu \chi$:
\begin{equation} \label{dea}
	S_{(\chi)}=\int d^4x \sqrt{-g}\,(\partial_{\mu}\chi)_{;\nu}T^{\mu\nu}_{(\chi)}\,.
\end{equation}
The variation with respect to $\chi$ gives a covariant conservation of a current $f^\mu$
\begin{equation} \label{force} 
	\nabla_{\mu}T_{\left(\chi\right)}^{\mu\nu}=f^\nu\,,\quad \nabla_{\nu}f^\nu=0\,,
\end{equation}
which it is the source of the stress energy-momentum tensor. Equation (\ref{force}) has a close correspondence to  (\ref{chitensor}). By taking variations with respect to $\chi_{\mu}$, we obtain 4 equations of motion which correspond to a  covariant conservation of the energy-momentum tensor $\nabla_\mu T^{\mu\nu}_{(\chi)}=0$. By changing the 4 vector to a gradient of a scalar $\partial_{\mu}\chi$, we change the conservation of energy-momentum tensor to an asymptotic conservation of energy-momentum tensor (\ref{force}) which corresponds to a conservation of a current $\nabla_{\nu}f^\nu=0$.  From a variation of the action with respect to the metric, we get a conserved stress energy tensor $T^{\mu\nu}_{(G)}$:
\begin{equation}
	T^{\mu\nu}_{(G)}=\frac{1}{\sqrt{-g}}\frac{\delta(\sqrt{-g}\mathcal{L}_M)}{\delta g^{\mu\nu}}\,,\quad \nabla_{\mu}T_{\left(G\right)}^{\mu\nu}=0\,.
\end{equation}

By considering $T^{\mu\nu}_{(\chi)}$ being equal to $\mathcal{L}_{1}g^{\mu\nu}$, the   
original measure $\Phi$ is modified to a Galileon measure $\Phi(\chi)= \partial_\mu (\sqrt{-g} g^{\mu\nu} \partial_\nu \chi) $, and the action \eqref{dea} gets the following form
\begin{equation} \label{GMT}
	S_{(\chi)}=\int d^4x \Phi(\chi) \mathcal{L}_{1}
\end{equation}
Here if we take variation with respect to the scalar $\chi$, the equation of motion gives $\Box \mathcal{L}_{1}=0$. This idea was also used in the context of string theory in \cite{Vulfs:2017dhl,Vulfs:2017ntw}.
 
	\section{The Action}\label{sec:1}
Let us start with the following two-measure action
\begin{equation}\label{action}
S=\frac{1}{2\kappa^2}\int d^{4}x\sqrt{-g}R+\int d^4x \sqrt{-g}\,(\Lambda(\phi,X)+V_1(\phi)) + \,\int d^{4}x \Phi(\chi) \, \Lambda(\phi,X)\,,
\end{equation} 
where $;$ represents covariant derivative with respect to the Levi-Civita connection and $\phi$ is a scalar field. The first two terms in the above action represents standard $k$-essence theories whereas the last term has another contribution with a different Galilean measure.

Then the variation with respect to the scalar $\chi$ gives $\Box \Lambda (\phi,X)=0$, which for a cosmological solution leads to an interactive unified DE/DM scenario \cite{G1,Benisty:2017eqh,Benisty:2017lmt}. The second term on this action depends on $\Lambda(\phi,X)$ which is a function of a scalar field $\phi$ and a kinetic term  \begin{equation}
X=-\frac{1}{2}\epsilon\,\partial_\mu\phi\partial^\mu\phi
\end{equation} 
that contains any $k$-essence theory. If $\epsilon=+1$, the scalar field $\phi$ represents a canonical scalar field, whereas when $\epsilon=-1$ represents a phantom scalar field. The third term in the action \eqref{action} also depends on an energy-momentum tensor $T_{\left(\chi\right)}^{\mu\nu}$ that couples to the vector field and it is assumed to be independent of it. In \cite{G1}, the authors studied the specific case where the function $\Lambda(\phi,X)$ is defined as follows
\begin{equation}
\Lambda(\phi,X)=K-V_2(\phi)=-\frac{1}{2}\epsilon\,\partial_\mu\phi\partial^\mu\phi-V_2(\phi)\,,\label{Lambda}
\end{equation}
where $V_2(\phi)$ is an energy potential, which in general is different than the potential $V_1(\phi)$. Note that the potentials are coupled with different measures. In \cite{G1}, the special case where $V_1(\phi)=V_2(\phi)=0$ was studied.

Variations of the action \eqref{action} with respect to the metric gives us the following field equations
\begin{equation} \label{fieldeq}
\begin{split}
G_{\mu\nu}=g_{\mu\nu}(\Lambda+\chi^{\lambda}\Lambda_{,\lambda})-j_\mu\phi_{,\nu} +\chi_{\mu}\Lambda_{,\nu}+\chi_{\nu}\Lambda_{,\mu}-g_{\mu\nu} V_1(\phi)\,,
\end{split}
\end{equation}
where we have assumed that commas denote differentiation, $\kappa^2=1$ and the vector field is equal to the gradient of the scalar field $\chi$ that appears in the Galileon measure, namely
\begin{equation}
\chi_\mu=\partial_\mu\chi\,.
\end{equation}
From a variation with respect to the scalar $\phi$ we obtain a non-conserved current, which is given by
\begin{equation} \label{current}
j_\alpha=2( \chi^\lambda_{;\lambda }+1)\phi_{,\alpha}\,.
\end{equation}
If we vary the action \eqref{action} with respect to the scalar field $\phi$ and the vector field $\chi_{\mu}$, we respectively~get 
\begin{eqnarray}
\frac{\epsilon}{2}\nabla_{\alpha}j^{\alpha}&=&\frac{dV_1(\phi)}{d\phi}+\frac{dV_2(\phi)}{d\phi}\Big(\chi^{\lambda}_{;\lambda}+1\Big)\,,\label{eqphi}\\
\square \Lambda&=&\square(K-V_2(\phi))=0\,.\label{eqLambda}
\end{eqnarray}
Here, $\square=\nabla_{\alpha}\nabla^{\alpha}$ is the d'Alambertian. In the next section, a spherically symmetric space-time of this model will be introduced in order to then analyse the special case of asymmetric wormholes. 
\section{Spherically symmetric space-time}\label{sec:2}
\subsection{General equations}
Let us start with the most general spherically symmetric space-time metric given by
\begin{eqnarray}
ds^2=-A(r)dt^2+B(r)dr^2+C(r)d\Omega^2\,,
\end{eqnarray}
where $A(r),B(r)$ and $C(r)$ are the metric coefficients which depend on the radial coordinate and $d\Omega^2=d\theta^2+\sin^2\theta d\phi^2$. In this space-time, the field equations \eqref{fieldeq} become
\begin{eqnarray}
&&\frac{4 B C C''-2 C B' C'-B C'{}^2-4 B^2 C}{4 B^2 C^2}+\frac{\epsilon  \phi'{}^2}{2 B}+V_1(\phi)+V_2(\phi)+ \frac{\phi'}{2B^3}\left(\epsilon B' \phi'-2 B^2 V_2'(\phi)-2 \epsilon  B \phi''\right)\chi_{r}=0\,,\label{1} \\
&&\frac{2 C A' C'+A C'^2-4ABC}{4 A B C^2}-\frac{\epsilon  \phi '^2}{2 B}+V_1(\phi)+V_2(\phi)-\frac{\epsilon  \chi'_{r} \phi '^2}{B^2}+ \frac{\phi'}{2 A B^2 C}\Big(2 A \left(B C V_2'(\phi)-\epsilon  (C \phi')'\right)-\epsilon  C A' \phi '\Big)\chi_{r}=0\,,\nonumber \\ \label{2}\\
&&\frac{A C \left(B A' C'+C \left(2 B A''-A' B'\right)\right)-B C^2 A'^2-A^2 C B' C'+2 A^2 B C C''-A^2 B C'^2}{4 A^2 B^2 C^2}+\frac{\epsilon  \phi '^2}{2 B}+V_1(\phi)+V_2(\phi)\nonumber\\
&&\ \ \ \ \ \ \ \ \ \ \ \ \ \ \ \ \ \ \ \ \ \ \ \ \ \ \ \ \ \ \ \ \ \ \ \ \ \ \ \ \ \ \ \ \ \ \ \ \ \ \ \ \ \ \ \ \ \ \ \ \ \ \ \ \ \ \ \ \ \ \ \ \ \ \ \ \ \ \ \ + \frac{\phi'}{2 B^3}\left(\epsilon  B' \phi'-2 B^2 V_2'(\phi)-2 \epsilon  B \phi ''\right)\chi_{r}=0\,.\label{3}
\end{eqnarray}
Here, prime denotes differentiation with respect to the radial coordinate $r$ and $\chi_\mu = (0,\chi',0,0):=(0,\chi_{r},0,0)$. Clearly, when $\chi_r=0$, one recovers standard scalar-tensor theory. The modified Klein-Gordon Equation \eqref{eqphi} becomes
\begin{eqnarray}
&&-\frac{\epsilon  A' \phi'}{2 A B}+\frac{\epsilon  B' \phi'}{2 B^2}+\frac{dV_1(\phi)}{d\phi}+\frac{dV_2(\phi)}{d\phi}-\frac{\epsilon  \left(C' \phi'+C \phi''\right)}{B C}-\frac{1}{B^3}\left[\epsilon  B \left(\frac{A' \phi'}{A}+\frac{2 C' \phi'}{C}+\phi''\right)-2 \epsilon  B' \phi'-\frac{dV_2(\phi)}{d\phi}B^2 \right]\chi'_r\nonumber\\
&&-\frac{\epsilon  \chi''_r\phi'}{B^2}+\frac{\chi_r}{4 A^2 B^4 C} \Big[\epsilon  B^2 C A'^2 \phi'+2 A B \Big\{2 \epsilon  C A' B' \phi'+\frac{d V_2(\phi)}{d\phi}B^2 C A' -\epsilon  B \left(C A'' \phi'+A' \left(2 C' \phi'+C \phi''\right)\right)\Big\}\nonumber\\
&&-A^2 \Big\{2 B^2 \left(\frac{dV_2(\phi)}{d\phi}C B'+2 \epsilon  \left(C'' \phi'+C' \phi''\right)\right)+5 \epsilon  C B'^2 \phi'-2 \epsilon  B \left(C B'' \phi'+B' \left(4 C' \phi'+C \phi''\right)\right)-4\frac{dV_2(\phi)}{d\phi} B^3 C' \Big\}\Big]=0\,,\nonumber \\ \label{klein2}
\end{eqnarray}
and the constraint \eqref{eqLambda} gives us
\begin{eqnarray}
\frac{d}{dr}\Big[\frac{\sqrt{A/B} C \phi' \left(\epsilon  B' \phi'-2 B^2 V_2'(\phi)-2 \epsilon  B \phi''\right)}{B}\Big]=0\,,\label{constraint}
\end{eqnarray}
which can be directly integrated yielding
\begin{eqnarray}
\frac{\sqrt{A/B} C \phi' \left(\epsilon  B' \phi'-2 B^2 V_2'(\phi)-2 \epsilon  B \phi''\right)}{B}=C_2\,,
\end{eqnarray}
where $C_2$ is an integration constant. There are four independent equations since the modified Klein-Gordon Equation \eqref{klein2} can be also obtained by using \eqref{1}--\eqref{3} and \eqref{constraint}. The constraint \eqref{constraint} is an additional equation that does not appear in standard $k$-essence theory. This equation comes directly by assuming that the vector field is a divergence of a scalar field. Note that if one subtracts~\eqref{1} with \eqref{3}, one gets 
\begin{eqnarray}
\frac{A''}{A B}-\frac{A' B'}{2 A B^2}+\frac{A' C'}{2 A B C}-\frac{A'^2}{2 A^2 B}+\frac{B' C'}{2 B^2 C}-\frac{C''}{B C}+\frac{2}{C}=0\,,\label{eqeasy1}
\end{eqnarray}
which is an equation which does not depend on the scalar fields. Moreover, this equation is valid for any $k$-essence theory as it pointed out in \cite{Bronnikov:2005gm}. The latter comes from the fact that in those theories $T_{t}^{t}=T_{\theta}^{\theta}$ and then all the contribution coming from the scalar field disappears.
\subsection{Asymmetric Wormholes triggering linear potentials describing galaxy halos}
In this section, we will choose that the metric coefficients are related as follows
\begin{eqnarray}
B(r)=\frac{1}{A(r)}\,.\label{ddd}
\end{eqnarray} 
It should be noted that one can define a new radial coordinate (see Equation~(2.4) in \cite{wheeler1995horizons}) and rewrite the metric only with two independent functions. Hence, choosing \eqref{ddd} is not an assumption, it is just a gauge choice \cite{wheeler1995horizons}.

By replacing the above equation into \eqref{eqeasy1}, one gets
\begin{eqnarray}
\frac{d}{dr}\Big(AC'-A'C\Big)=-\frac{d}{dr}\Big[C^2\frac{d}{dr}\Big(\frac{A}{C}\Big)\Big]=2\,,
\end{eqnarray}
which is the same equation reported in~\cite{Bronnikov:2005gm}. Then, the global geometric structure would be the same as described in the latter mentioned paper. Then, one can easily integrate once to obtain
\begin{eqnarray}\label{AC}
\frac{d}{dr}\Big(\frac{A}{C}\Big)=\frac{2(r_0-r)}{C^2}\,,
\end{eqnarray} 
where $r_0$ is an integration constant. This result is generic for any scalar field $\phi$, $\chi_r$ and also for any energy potential $V_1(\phi)$ and $V_2(\phi)$. 

 Let us further assume an asymmetric wormhole geometry with the  following metric coefficient~function
 \begin{equation}\label{C}
 C(r)=r^2+b^2+a r\,,
 \end{equation}
where $b$ and $a$ are the wormhole parameters. The parameter $a$ measures the asymmetry of the wormhole throat under $r\rightarrow -r$ and therefore the asymmetry between the two sides of the wormhole, although as we  will see, there is another radius in the wormhole with respect to which one can look at its asymmetry, which we will argue is more relevant, which is the centre of gravity sphere (since it correspond to a certain radius, but the angles are arbitrary)  . Then, one can directly solve \eqref{AC} yielding
\begin{eqnarray}\label{Ais}
A(r)=A_0 \left(a r+b^2+r^2\right)+\frac{1}{c^2}\Big[\frac{4 (a+2 r_0) \left(a r+b^2+r^2\right) \arctan\left(\frac{a+2 r}{c}\right)}{c}+2 \left(a (r+r_0)+2 b^2+2 r r_0\right)\Big]\,,
\end{eqnarray}
  where $A_0$ is an integration constant and for simplicity we have defined $c=\sqrt{4b^2-a^2}$. $\pi c$ measures the circumferential radius of the wormhole at its neck. The parameters must satisfy $-2 b< a< 2 b$ which also ensures that the zeros of the equation $C(r)=0$ will be imaginary making that the wormhole exists. In general, the space-time can be either a wormhole or a black hole depending on the parameters. Let us here emphasise again that the above solutions are very general since they do not depend on the matter/scalar field chosen. Equation~\eqref{eqeasy1} is a purely geometric equation which arises directly from the field equations (by subtracting \eqref{1} with \eqref{3}). The latter equation is valid for any $k$-essence theory and also for any two-measure theory, independently of the source.
 
\begin{figure}[H]
	\centering
	\includegraphics[width=0.6\textwidth]{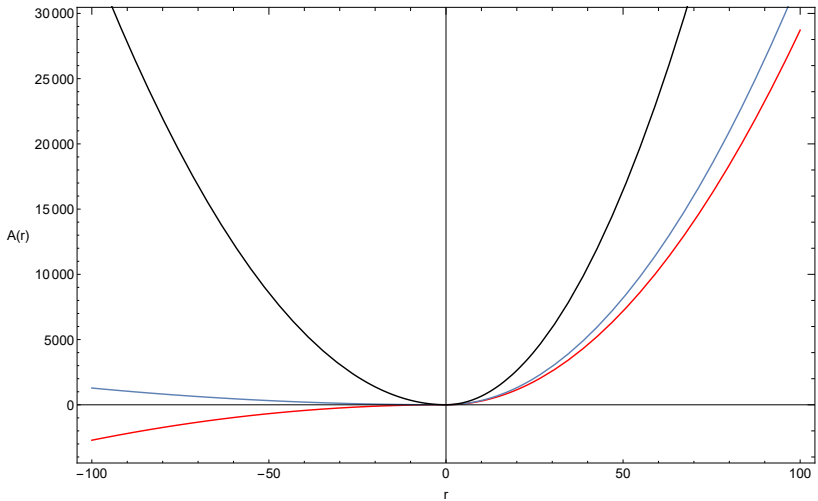}
	\caption{Plots of the coefficient $A(r)$ versus $r$ for different values of $A_0$. The red, blue and black solid lines represent $A_0=1.3$, $A_0=1.7$ and $A_0=5$ respectively. We have chosen the parameters $a=0.001$ and $b=r_0=1$ and $A_0=1$. Negative values of $r$ are also plotted which represent the other universe.}
	\label{fig1}
\end{figure}

 Let us study some special limit cases for our model. If one assumes that $2r+a\ll c$ and $r\ll a$,  we~can expand the metric coefficient up to second power-law orders in $r$, obtaining
 \begin{eqnarray}
A(r)\approx 1+A_0b^2+\frac{a c (\lambda -A_0)}{\pi }+\frac{r (\pi a A_0 -2 A_0 c+2 c \lambda )}{\pi }+A_0r^2+\mathcal{O}(r^3)\,.
 \end{eqnarray} 
Here, we have used the expansion $\arctan(x)\approx x $ for $x\ll 1$ and for simplicity we have introduced the following constant
 \begin{equation}\label{lambda}
\lambda=A_0+\frac{2 \pi  (a+2 r_0)}{c^3}\,.
\end{equation}
Now, let us explore the limit case at very large scales. In this case, one can assume that $2r+a\gg c$ and $r\gg a$ and then one can expand the metric function \eqref{Ais} as follows
 \begin{eqnarray}\label{AAA}
A(r)\approx 1+b^2 \lambda  +\lambda r^2+a\lambda r +\frac{c^3 (A_0-\lambda )}{6\pi r}+\mathcal{O}\Big(\frac{1}{r^2}\Big)\,,
 \end{eqnarray}
where we have used the expansion $\arctan(x)\approx \pi/2-1/x$ for $x\gg 1$. Then, $\lambda$ can be interpreted as a cosmological constant and the term $\lambda r^2$ will be important at cosmological scales. At galactic scales, the leading terms will be proportional to linear potential and inverse potential, namely
\begin{eqnarray}
a\lambda  r -\frac{c^3 (\lambda -A_0)}{6\pi r}\,.
\end{eqnarray}
One can directly see from Eq.~\eqref{AAA} that the space-time could be a wormhole if 
\begin{equation}
A_0 > \frac{2 \pi  (a+2 r_0)}{c^3}\,.\label{ineq}
\end{equation}
This condition tells us that at $r\rightarrow \pm\infty$, the metric coefficient $A(r)$ will have the same sign, if in these two asymptotic limits  $r\rightarrow \pm\infty$ the metric coefficient $A(r)$  has different signs, we are guaranteed that the metric coefficient $A(r)$ goes through zero at some point, and therefore we have a black hole solution, but if at those two limits  $A(r)$ has the same sign, the solution could be a wormhole. This is a necessary but may be not sufficient condition . As  examples for this, Figure~\ref{fig1} shows the metric coefficient $A(r)$ for some specific values of the parameters which give  asymmetric wormhole for some choices of the parameters, or alternatively a black hole. Three different cases are displayed: The red line shows the case where the inequality \eqref{ineq} does not hold, therefore, it does not describe a wormhole. In this case, the function is describing two black holes with an asymptotically de-Sitter space for the first one (positive $r$) and an anti de-Sitter space on the other side (negative $r$). On the contrary, when \eqref{ineq} holds, one has that $A(r)$ is describing an asymptotically de-Sitter space of one side of the wormhole (positive $r$) and also a de-Sitter space on the other side (negative $r$). This case is shown in blue and black lines which clearly, represents an asymmetric wormhole.

As we will see in the next section, to see the physical implications one should express these results in terms of the centre of gravity of the wormhole. Calculations in classical mechanics are often simplified when laws are formulated with respect to the centre of mass. In this case, the centre of mass is a hypothetical sphere where entire mass of an object may be assumed to be concentrated to visualise its motion. In other words, the centre of mass is the single radius equivalent for the  application of Newton's laws similar to the case of ordinary non-relativistic mechanics.
We now go on to define the analogous of this concept for the case of an asymmetric wormhole.

\section{Centre of gravity coordinates}

The wormhole has two special radii. The first one is the neck, where $C(r)$ is a minimum, and from (\ref{C}) we obtain that the radial coordinate for this is $r_m=-a/2$. The second one is the centre of gravity radius, or the equilibrium radius, where the Newtonian gravitational force vanishes. In general, when one considers an extended object, one defines a worldline
on the basis of the centre of mass, as discussed by Pound \cite{Pound}. So here also, for analysing the behaviour with respect to this special location, we have to use the coordinates of the centre of gravity by considering a shift
\begin{equation}\label{sc}
r = r' + \Delta\,.
\end{equation}
The coordinates of the centre of mass are a type of collective coordinates used when one is
dealing with extended objects, and therefore since a wormhole  is an extended object, it is natural to use them. So if the wormhole interacts with  another wormhole or with a
point particle, the use  of coordinates that vanish at  the centre of
gravity is preferable since the centre of gravity coordinate truly describes the collective motion of the extended object. Then, other coordinate choices will not be so physically~correct.

Demanding that for small $r'$, $A(r')$ does not contain linear terms in $r'$ and inserting (\ref{sc}) into ($\ref{Ais}$), we obtain that the linear term of $A(r')$ is cancelled for the following choice of $\Delta$,
\begin{equation}
\Delta=\frac{c}{\pi}-\frac{a}{2}-\frac{c\lambda}{A_0 \pi}\,.
\end{equation}
Now, by expressing the small $r'$ limit in terms of the centre of gravity coordinates, we find
\begin{equation}
A(r')= 1+ A_0 b^2 +\frac{ac(\lambda - A_0)}{\pi}+ A_0 r'^2 - A_0 \Delta^2\,.
\end{equation}
where we see that the linear terms are now cancelled. We can see that for positive values of $A_0$ that the Newtonian potential produces attraction towards the centre of gravity point $r'= 0$ for small $r'$, so the radius $r'= 0$ is indeed the radius towards where test particles are attracted to. This is therefore the centre of gravity radius. Notice that $r'= 0$, with whatever constant angles we choose, represent a geodesic motion. To ensure that the metric has the correct signature for small $r'$, we require also that $$1+ A_0 b^2 +\frac{ac(\lambda - A_0)}{\pi} - A_0 \Delta^2 > 0.$$

In the case where $r' \gg c(\frac{1}{2}-\frac{c}{\pi}+\frac{\lambda}{A_0\pi})$, we obtain that the $00$ component of the metric becomes,
\begin{equation}
A(r')= (1+ \lambda b^2+ a\lambda \Delta + \lambda \Delta^2) -\frac{2\lambda c(\lambda - A_0)}{A_0}r' + \lambda r'^2 - \frac{ c^3(\lambda - A_0)}{6 \pi r'}\,.
\end{equation}
In the $00$ component of the metric, the coefficient of the $1/r'$ term equals to $-2M$, where $M$ is the mass of the wormhole. Therefore one has that 
\begin{equation}
M=\frac{c^3(\lambda - A_0)}{12\pi}
\end{equation}
and from Eq.~(\ref{lambda}) we can get the dependence of the mass in terms of $a$ and $r_0$
\begin{equation}
M=\frac{a+2r_0}{6}\,.
\end{equation}
The linear term of $A(r')$ for large $r'$ can be expressed in terms of $M$, obtaining, 
 \begin{equation}
A(r')= (1+ \lambda b^2 + a\lambda \Delta + \lambda \Delta^2) -\frac{24\lambda M }{A_0 c^2}r' + \lambda r'^2 - \frac{ 2M}{ r'}\,.\label{41}
\end{equation}
The coefficient that multiplies $r'^2$ for large $|r'|$ identified the values of the cosmological constant at the two sides of the wormhole. Because of: 
\begin{equation}
\lambda_\pm = A_0 \pm \frac{2\pi(a+r_0)}{c^3}=-\frac{\Lambda_\pm}{3}\,.
\end{equation} 
The discontinuity of the cosmological constant between the two asymptotic sides of the wormhole~is:
\begin{equation}
 \Lambda_+- \Lambda_- = -\frac{72M}{c^3}\,.
\end{equation}
A combination of a Newtonian potential with linear $r'$ and inverse of $r'$  can provide an explanation for flat rotation curves without introducing dark matter. Moreover, in four different recent studies~\cite{OBrien:2017bwr,Mannheim:2010xw,OBrien:2011vks,Mannheim:2010ti}, the authors found that this combination fits well with 110 different spiral galaxies and also 25 dwarf galaxies. Furthermore, there is only one free parameter for each galaxy, viz., the mass to light ratio of each galaxy, and yet with no flexibility the	fit capture the essence of the data. Hence, invoking the presence of dark matter may be nothing more than an attempt to describe global effects in purely local galactic terms. On the contrary, the standard NFW profile or other dark matter profiles have more parameters for each galaxy. In order fit the same 138 galaxies data studied~in~\cite{Mannheim:2010ti} as fitting parameters, one needs two additional free parameters for each galactic halo. This gives us 276 free parameters to fit for a dark matter profile approach. Thus, these kinds of potentials are physically very well motivated and describe galaxy rotation curves in good agreement with observations. An interesting point arises here. The wormhole parameters $a$, $r_0$ and $b$ appear in the constants which are related to the flat rotation curves described in \cite{Mannheim:2016lnx,Mannheim:2010ti}. Thus, the asymmetric wormhole is acting as a trigger of a dark matter behaviour. Notice that the linear term in the Newtonian potential $r'(4\Lambda_+ M)/(A_0 c^2)$ is proportional to the mass of the wormhole  $M$. 
For positive values of $M, A_0, \Lambda_+$ we obtain that the Newtonian potential  produces  an attractive force. Notice that using the coordinates of the centre of gravity $r'$, the linear term and the $1/r'$ term are both proportional to the mass $M$, as is the discontinuity in the cosmological constant across the wormhole. So the mass appears as the source of gravitational attraction, at both large and small distances. As well us being the source of the discontinuity of the cosmological constant across the wormhole. 

Notice that the linear potential that appears in \eqref{41} is in fact proportional to $\lambda$ and the mass of the gravitating objects. Since $\lambda$ is connected with the Hubble constant $H_0$, we see that there is a connection between the linear potential governing the dark matter sector and the Hubble constant governing the acceleration of the universe. This seems to be related to Milgrom's idea (MOND) \cite{Milgrom:1992hr}, who advocates a relation between the minimal acceleration $a_0$ and the Hubble constant $H_0$.

This solution and also this interpretation would be also valid for any other $k$-essence theory. Then, one can say that if ones assumes an asymmetric wormhole geometry, the potential could describe dark matter and dark energy in a unified form.

Let us finish this section by noticing that according to~\cite{OBrien:2017bwr}, it was shown that by having linear and inverse potential terms, one can derive from first principles, the Tully-Fisher relation.

\section{The behaviour for the scalar potentials}
\begin{figure}[H]
	\centering
	\includegraphics[width=0.6\textwidth]{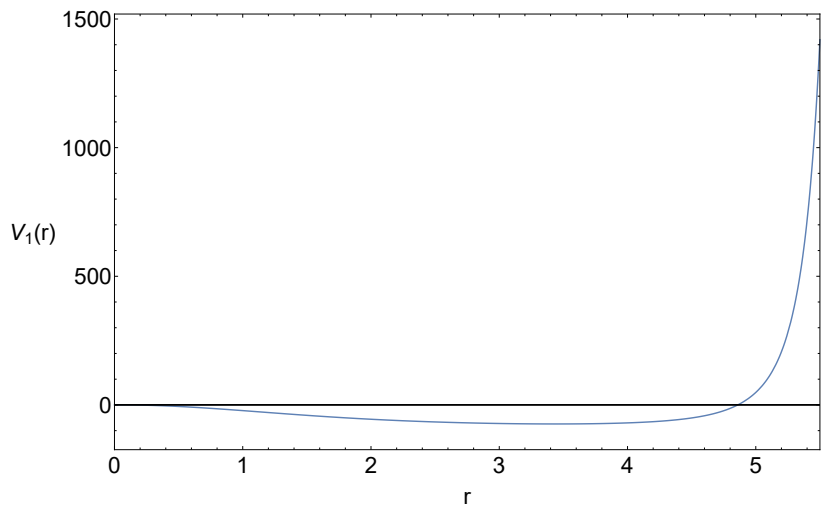}
	\caption{Plot of the potential $V_1(r)$ versus $r$ for $\phi_0=1$, $b=2.5$, $a=0.001$, $\lambda=0.00001$, $C_2=0.1$, $V_0=100$  and $A_0=-1$.}
	\label{fig2}
\end{figure}

In this section, we will assume a canonical scalar field ($\epsilon=+1$). In order to find solutions for our specific diffusive two measures theory, one needs to impose an additional ansatz since we have more variables than remaining equations. As an example and completeness, let us assume that the scalar field behaves as
 \begin{equation}
 \phi=\phi_0\arctan\Big(\frac{a+2r}{c}\Big)\,.\label{ansatz}
 \end{equation}
Here, $\phi_0$ is a constant. Then, by replacing this form into \eqref{constraint}, one can easily find that the potential takes the following form
 \begin{eqnarray}
 V_2(\phi)&=&\frac{1}{8 \pi  c^2 \phi_0}\Big[-2 \phi_0^2 \cos \left(\frac{2 \phi}{\phi_0}\right) \left(2 c^2 \phi  (\lambda -A_0)+\pi  \phi_0 \left(A_0 c^2+4\right)\right)+c \Big(-4 \phi   \left(c \phi_0^2 (\lambda -A_0)+4 \pi  C_2\right)\nonumber\\
 &&+2 c \phi_0^3 (A_0-\lambda ) \sin \left(\frac{2  \phi }{\phi_0}\right)+c \phi_0^3 (A_0-\lambda ) \sin \left(\frac{4 \phi }{\phi_0}\right)\Big)-2 \pi  \phi_0^3 \cos \left(\frac{4 \phi }{\phi_0}\right)\Big]+V_0\,,
 \end{eqnarray}
where $V_0$ is an integration constant. Now, the scalar field $\chi_r$ can be directly solved by using \eqref{1}; however, it depends implicitly on $V_1(\phi)$. In order to find $V_1(\phi)$ one needs to solve the remaining Eq.~\eqref{2}. This equation is an ordinary first order equation which in principle has a solution but analytically, it can be easily solved. One can then solve that equation numerically to see the behaviour of the potential $V_1(\phi)$. As one can see form Figure~\ref{fig2}, the potential $V_1(\phi)$ behaves with well defined asymptotic properties. We present this to show the existence of solutions. A variety of other solutions, starting from an ansatz different than \eqref{ansatz} could be explore also. As it can be seen from Equation~\eqref{action}, the potential $V_1(\phi)$ is coupled with the standard volume measure $\sqrt{-g}$ and acts like a vacuum energy potential (as can be seen in Eq.~\eqref{fieldeq}). In contrast, $V_2(\phi)$ is coupled with the modified volume measure $\Phi(\chi)$ and gives an equation on how the function $\Lambda(\phi,X)$ evolves in space-time. In our model, $V_2(\phi)$ was easily found analytically but $V_1(\phi)$ only was found numerically. Note that these potentials act like matter supporting the wormhole and they are not related to the metric which gives the gravitational potential of the wormhole.

\section{Conclusions}
   In this paper we have explored wormholes solutions in a particular DE/DM unified model described in \cite{G1,Benisty:2017lmt}. In this case and for asymmetric wormholes, we have found that the asymmetry between the two universes connected through a wormhole induces a linear term in the gravitational potential, and have calculated the coefficient of these linear term in the coordinates of the centre of gravity of the wormhole. These coordinates are expected to be the most suitable ones if we are  interested in the collective motion of the wormhole as is the coordinates of the centre of gravity in non-relativistic mechanics. As discussed in \cite{Mannheim:2016lnx}, these linear gravitational potentials can be used to explain the behaviour of galactic rotation curves. The idea that the massive object at the centre of our galaxy is a wormhole rather than a black hole has been discussed
  together with some possible observational consequences related to the effect of this on the geodesics produced by this object, if it is indeed a wormhole \cite{Zilong} . These effects are indeed even more accurate in the case of the solutions discussed in this paper  due to the generation of the linear potentials, which as have argued, could represent effects of dark matter. Let us stress here that even though we focused our study in wormhole geometries, the coefficients of the space-time \eqref{C} and \eqref{Ais} can also describe black hole solutions with a linear potential. This can be directly seen from Figure~\ref{fig1} since if Equation~\eqref{ineq} is not valid, the geometry will be a black hole. This happens since $A(r)$ crosses zero showing the horizon at this point. Hence, effectively, the geometry can describe either a wormhole or a black hole in the centre of a galaxy. 
\begin{acknowledgments}
S.B. is supported by the Comisi{\'o}n Nacional de Investigaci{\'o}n Cient{\'{\i}}fica y Tecnol{\'o}gica (Becas Chile Grant No.~72150066). The authors would also like
to acknowledge networking support by the COST Action GWverse CA16104. E.G. is grateful for many interesting discussions of the subject of this paper with Phillip Mannheim and Thom Curtright during the Miami 2017 conference and latter on during a visit to the Physics Department of the University of Miami and all of us are thankful also for additional communications in the course of writing this paper with Phillip Mannheim. We also acknowledge Leor Barak for conversations concerning the need to refer to the centre of mass during the first global event of the COST Action GWverse CA16104.
\end{acknowledgments}


\end{document}